# Approximate solution of the Duffin–Kemmer–Petiau equation for a vector Yukawa potential with arbitrary total angular momenta

M. Hamzavi[1*], S. M. Ikhdair[2**]

[1]*Department of Basic Sciences, Shahrood Branch, Islamic Azad University, Shahrood, Iran*

[2]*Physics Department, Near East University, 922022 Nicosia, North Cyprus, Mersin 10, Turkey*

[*]*Corresponding author: Tel.:+98 273 3395270, fax: +98 273 3395270*

[*]Email: majid.hamzavi@gmail.com

[**]Email: sikhdair@neu.edu.tr

## Abstract

The usual approximation scheme is used to study the solution of the Duffin–Kemmer–Petiau (DKP) equation for a vector Yukawa potential in the framework of the parametric Nikiforov-Uvarov (NU) method. The approximate energy eigenvalue equation and the corresponding wave function spinor components are calculated for any total angular momentum $J$ in closed form. Further, the exact energy equation and wave function spinor components are also given for the $J=0$ case. A set of parameter values is used to obtain the numerical values for the energy states with various values of quantum levels $(n, J)$.

**Keywords:** DKP equation; Nikiforov-Uvarov method; Yukawa potential.

**PACS:** 03.65.Ca, 03.65.Nk, 03.65.Pm, 02.30.Em

## 1. Introduction

The first-order DKP formalism has been used to analyze relativistic interactions of spin-0 and spin-1 hadrons with nuclei as an alternative to their conventional second-order Klein-Gordon (KG) and Proca equations [1-5]. The DKP equation is a direct generalization to the Dirac particles of integer spin in which one replaces the gamma matrices by beta metrics but verifying a more complicated algebra as DKP algebra [6-14]. Fainberg and Pimentel presented a strict proof of equivalence between DKP and KG theories for physical S-matrix elements in the case of charged scalar particles interacting in minimal way with an external or quantized electromagnetic field [15-16]. Boutabia-Chéraitia and Boudjedaa solved the DKP equation in the presence of Woods–Saxon potential for spin 1 and spin 0 and they also deduced the transmission and reflection coefficients [17]. Kulikov et al. offered a new oscillator model with different form of the nonminimal substitution within the framework of the DKP equation [18]. Yaşuk et al. presented an application of the relativistic DKP equation in the presence of a vector deformed Hulthén potential for spin zero particles by using the Nikiforov-Uvarov (NU) method [19]. Boztosun et al. presented a simple exact analytical solution of the relativistic DKP equation within the framework of the asymptotic iteration method and determined exact bound state energy eigenvalues and corresponding eigenfunctions for the relativistic harmonic oscillator and the Coulomb potentials [20]. Kasri and Chetouani determined the bound state energy eigenvalues for the relativistic DKP oscillator and DKP Coulomb potentials by using an exact quantization rule [21]. de Castro explored the problem of spin-0 and spin-1 bosons subject to a general mixing of minimal and nonminimal vector cusp potentials in a unified way in the context of the DKP theory [22]. Chargui et al. solved the DKP equation with a pseudoscalar linear plus Coulomb-like potential in a two-dimensional space–time [23]. Very recently, Hamzavi and Ikhdair obtained solution of the DKP equation for a vector deformed Woods-Saxon potential by using the Pekeris approximation [24].

In this work, we study the Yukawa potential or static screening Coulomb potential (SSCP) which is given by

$$V(r) = -V_0 \frac{e^{-ar}}{r}, \tag{1}$$

where $V_0 = \alpha Z$, $\alpha = 1/137.037$ is the fine-structure constant with $Z$ is the atomic number and $a$ is the screening parameter. This potential is often used to compute bound-state normalizations and energy levels of neutral atoms [25-28] which have been studied over the past years. It is well known that SSCP yields reasonable results only for the innermost states when $Z$ is large. However, for the outermost and middle atomic states, it gives rather poor results. The bound-state energies of the SSCP with $Z = 1$ have been studied in the light of the shifted large-$N$ expansion method [29]. Ikhdair and Sever investigated energy levels of neutral atoms by applying an alternative perturbative scheme in solving the Schrödinger equation for the Yukawa potential model with a modified screening parameter [28]. They studied bound states of the Hellmann potential, which represents the superposition of the attractive Coulomb potential $-a/r$ and the Yukawa potential $b\exp(-\delta r)/r$ of arbitrary strength $b$ and the screening parameter $\delta$ [30]. They also considered the bound states of the exponential cosine-screened Coulomb potential [31] and a more general exponential screened Coulomb (MGESC) potential [32].

Recently, Karakoc and Boztosun applied the asymptotic iteration method to solve the radial Schrödinger equation for the Yukawa type potentials [33]. Further, Gönül *et al* presented an approximation scheme to obtain analytical expressions for the bound state energies and wave functions of Yukawa like potentials [34]. Very recently, Ikhdair has approximately solved the relativistic Dirac equation with the Yukawa potential for any spin-orbit quantum number $\kappa$ in the presence of spin and pseudospin symmetry [35]. In addition, Liverts *et al* used the quasi-linearization method (QLM) for solving the Schrödinger equation with Yukawa potential [36].

Therefore, it would be interesting and important to solve the DKP equation for the Yukawa potential since it has been extensively used to describe the bound and scattering states of the above mentioned interesting deformed nuclear models. In this paper, we will study the DKP equation with a vector Yukawa potential for non-zero total angular momentum, i.e. $J \neq 0$. Under these conditions, the DKP equation has no exact solutions and we use the appropriate approximation.

This work is organized as follows: in section 2, the DKP formalism is given briefly and discussed under a vector potential. In section 3, the parametric generalization of the NU method is introduced. We solve the DKP equation with a vector Yukawa potential in section 4. We also obtain numerical energy results for any arbitrary $(n,J)$ states. The exact energy equation and wave function spinor components are also given for the $J=0$ case in section 4. Finally, section 5 is devoted to our conclusion.

## 2. A Brief Review to the DKP Formalism

The first order relativistic DKP equation for a free spin-0 or spin-1 particles of mass $m$ is

$$(i\beta^{\mu}\partial_{\mu}-m)\psi_{DKP}=0, \tag{2}$$

where $\beta^{\mu}$ ($\mu=0,1,2,3$) matrices satisfy the commutation relation

$$\beta^{\mu}\beta^{\nu}\beta^{\lambda}+\beta^{\lambda}\beta^{\nu}\beta^{\mu}=g^{\mu\nu}\beta^{\lambda}+g^{\nu\lambda}\beta^{\mu}, \tag{3}$$

which defines the so-called DKP algebra. The algebra generated by the $4\beta^N$ 's has three irreducible representations: a ten dimensional one that is related to $S=1$, a five dimensional one relevant for $S=0$ (spinless particles) and a one dimensional one which is trivial. In the spin-0 representation, $\beta^\mu$ are $5\times 5$ matrices defined as ($i=1,2,3$)

$$\beta^0=\begin{pmatrix}\theta & \tilde{0}\\ \bar{0}_T & 0\end{pmatrix},\qquad \beta^i=\begin{pmatrix}\tilde{0} & \rho^i\\ -\rho_T^i & 0\end{pmatrix}, \tag{4}$$

with $\tilde{0}$, $\bar{0}$, $0$ as $2\times 2$, $2\times 2$, $3\times 3$ zero matrices, respectively, and

$$\theta=\begin{pmatrix}0 & 1\\ 1 & 0\end{pmatrix},\ \rho^1=\begin{pmatrix}-1 & 0 & 0\\ 0 & 0 & 0\end{pmatrix},\ \rho^2=\begin{pmatrix}0 & -1 & 0\\ 0 & 0 & 0\end{pmatrix},\ \rho^3=\begin{pmatrix}0 & 0 & -1\\ 0 & 0 & 0\end{pmatrix}. \tag{5}$$

For spin one particle, $\beta^\mu$ are $10\times 10$ matrices given by

$$\beta^0=\begin{pmatrix}0 & \bar{0} & \bar{0} & \bar{0}\\ \bar{0}^T & 0 & I & I\\ \bar{0}^T & I & 0 & 0\\ \bar{0}^T & 0 & 0 & 0\end{pmatrix},\qquad \beta^i=\begin{pmatrix}0 & \bar{0} & e_i & \bar{0}\\ \bar{0}^T & 0 & 0 & -is_i\\ e_i^T & 0 & 0 & 0\\ \bar{0}^T & -is_i & 0 & 0\end{pmatrix}, \tag{6}$$

where $s_i$ are the usual $3\times 3$ spin one matrices

$$\bar{0}=\begin{pmatrix}0 & 0 & 0\end{pmatrix},\ e_1=\begin{pmatrix}1 & 0 & 0\end{pmatrix},\ e_2=\begin{pmatrix}0 & 1 & 0\end{pmatrix},\ e_3=\begin{pmatrix}0 & 0 & 1\end{pmatrix}, \tag{7}$$

$I$ and $0$ are the identity and zero matrices, respectively. While the dynamical state $\psi_{DKP}$ is a five component spinor for spin zero particles, it has ten component spinors for $S=1$ particles. The solution of the DKP equation for a particle in a central field needs consideration since earlier work [11]. It is convenient to recall some general properties of the solution of the DKP equation in a central interaction for spin zero particle. The central interaction consists of two parts: a Lorentz scalar $U_S$ and a time-like component of four-dimensional vector potential, $U_V$ depending only on $r$ [19,22]. The stationary states of the DKP particle, in units $\hbar=c=1$, in this case are determined by solving

$$\left(\vec{\beta}.\vec{p}+m+U_S+\beta^0U_V\right)\psi(\vec{r})=\beta^0E\psi(\vec{r}). \tag{8}$$

In the spin zero representation, the five component DKP spinor is

$$\psi(\vec{r})=\begin{pmatrix}\psi_{\text{upper}}\\ i\psi_{\text{lower}}\end{pmatrix}\ \text{ with }\ \psi_{\text{upper}}=\begin{pmatrix}\phi\\ \varphi\end{pmatrix}\ \text{ and }\ \psi_{\text{lower}}=\begin{pmatrix}A_1\\ A_2\\ A_3\end{pmatrix}, \tag{9}$$

so that for stationary states the DKP equation can be written as

$$(m+U_S)\phi=(E-U_V^0)\varphi+\vec{\nabla}.\vec{A}, \tag{10}$$

$$\vec{\nabla}\phi=(m+U_S)\vec{A}, \tag{11}$$

$$(m+U_S)\varphi=(E-U_V^0)\phi, \tag{12}$$

where $\vec{A}$ is the vector $(A_1,A_2,A_3)$. The five-component wave function $\psi$ is simultaneously an eigenfunction of $J^2$ and $J_3$

$$J^2\begin{pmatrix}\psi_{\text{upper}}\\ \psi_{\text{lower}}\end{pmatrix}=\begin{pmatrix}L^2\psi_{\text{upper}}\\ (L+S)\psi_{\text{lower}}\end{pmatrix}=J(J+1)\begin{pmatrix}\psi_{\text{upper}}\\ \psi_{\text{lower}}\end{pmatrix}, \tag{13}$$

$$J_3\begin{pmatrix}\psi_{\text{upper}}\\ \psi_{\text{lower}}\end{pmatrix}=\begin{pmatrix}L_3\psi_{\text{upper}}\\ (L_3+S_3)\psi_{\text{lower}}\end{pmatrix}=M\begin{pmatrix}\psi_{\text{upper}}\\ \psi_{\text{lower}}\end{pmatrix}, \tag{14}$$

where the total angular momentum $J=L+S$ which commutes with $\beta^0$, is a constant of the motion. The most general solution of Eq. (8) is

$$\psi_{JM}(r)=\begin{pmatrix}f_{nJ}(r)Y_{JM}(\Omega)\\ g_{nJ}(r)Y_{JM}(\Omega)\\ i\sum_L f_{nJL}(r)Y_{JL1}^M(\Omega)\end{pmatrix}, \tag{15}$$

where $Y_{JM}(\Omega)$, the spherical harmonics of order $J$, $Y_{JL1}^M(\Omega)$ are the normalized vector spherical harmonics and $f_{nJ}(r)$, $g_{nJ}(r)$ and $f_{nJL}(r)$ are radial wave functions. The insertion of $\psi_{JM}(r)$ defined in Eq. (15) into Eqs. (10), (11) and (12) by making use of the properties of vector spherical harmonics [2] yields the following set of first-order coupled relativistic differential radial equations

$$(E-U_V^0)F(r)=(m+U_S)G(r), \tag{16a}$$

$$\left(\frac{d}{dr}-\frac{J+1}{r}\right)F(r)=-\frac{1}{\alpha_J}(m+U_S)H_1(r), \tag{16b}$$

$$\left(\frac{d}{dr}+\frac{J}{r}\right)F(r)=-\frac{1}{\varsigma_J}(m+U_S)H_{-1}(r), \tag{16c}$$

$$\begin{aligned}-\alpha_J\left(\frac{d}{dr}+\frac{J+1}{r}\right)H_1(r)+\varsigma\left(\frac{d}{dr}-\frac{J}{r}\right)H_{-1}(r)\\ =(m+U_S)F(r)-(E-U_V^0)G(r),\end{aligned} \tag{16d}$$

where $\alpha_J = \sqrt{(J+1)/(2J+1)}$, $\varsigma_J = \sqrt{J/(2J+1)}$, $f_{nJ}(r) = F(r)/r$, $g_{nJ}(r) = G(r)/r$ and $h_{nJJ\pm1}(r) = H_{\pm1}(r)/r$. For the DKP equation, at the presence of vector potential and while the scalar potential is taken zero, the differential equations to be satisfied by the radial wave functions are

$$(E - U_V^0)F(r) = mG(r), \tag{17a}$$

$$\left(\frac{d}{dr} - \frac{J+1}{r}\right)F(r) = -\frac{1}{\alpha_J} mH_1(r), \tag{17b}$$

$$\left(\frac{d}{dr} + \frac{J}{r}\right)F(r) = -\frac{1}{\varsigma_J} mH_{-1}(r), \tag{17c}$$

$$\begin{aligned} -\alpha_J \left(\frac{d}{dr} + \frac{J+1}{r}\right)H_1(r) + \varsigma \left(\frac{d}{dr} - \frac{J}{r}\right)H_{-1}(r) \\ = mF(r) - (E - U_V^0)G(r). \end{aligned} \tag{17d}$$

Eliminating $G(r)$, $H_1(r)$ and $H_{-1}(r)$ in terms of $F(r)$, the following second-order differential equation is satisfied by the function $F(r)$,

$$\vec{O}_{\mathrm{KG}} F(r) = 0, \tag{18}$$

that represents the radial KG equation for a vector potential with $\vec{O}_{\mathrm{KG}}$ is the KG operator defined as

$$\vec{O}_{\mathrm{KG}} \rightarrow \vec{\nabla}^2 + \left(E - eV(r)\right)^2 - m^2,\ \vec{\nabla}^2 \rightarrow \frac{d^2}{dr^2} - \frac{J(J+1)}{r^2}. \tag{19}$$

It is remarkable to note that Eq. (18) is equivalent to Eq. (16) of Ref. [17]. Alternatively, we can rewrite (18) as the radial KG equation:

$$\left[\frac{d^2}{dr^2} - \frac{J(J+1)}{r^2} + \left(E - U_V^0\right)^2 - m^2\right]F(r) = 0,\ U_V^0 = -eU_0 \frac{e^{-ar}}{r}, \tag{20}$$

where $J(J+1)/r^2$, is the total angular momentum centrifugal term and we take $e = 1$ [19].

Here, we take the vector potential in Eq. (20) as the Yukawa potential (1). Therefore, the radial DKP equation for $F(r)$ reduces to

$$\left[\frac{d^2}{dr^2} - \frac{J(J+1)}{r^2} + \frac{U_0^2 e^{-2ar}}{r^2} + \frac{2EU_0 e^{-ar}}{r} + E^2 - m^2\right]F(r) = 0. \tag{21}$$

Since Eq. (21) does not admit exact analytical solution due to the presence of the strong singular centrifugal term $r^{-2}$, we resort to use a proper approximation to deal with this term. So, we employ the conventional approximation scheme introduced by Greene and Aldrich [37]:

$$\frac{1}{r^2} \approx 4a^2 \frac{e^{-2ar}}{(1-e^{-2ar})^2}, \tag{22a}$$

$$\frac{1}{r} \approx 2a \frac{e^{-ar}}{(1-e^{-2ar})}, \tag{22b}$$

which is valid only for a short-range potential, i.e., $ar \ll 1$ and used only to calculate the lowest energy states as mentioned in [38]. Therefore, to see the accuracy of our approximation, we plot the Yukawa potential (1) and its approximation [35,39]

$$V(r) = -2aU_0 \frac{e^{-2ar}}{1-e^{-2ar}}, \tag{23}$$

with parameters $U_0 = 1.0$ and $a = 0.01 fm^{-1}$, as shown in Figure 1. Thus, the approximate analytical solution of the DKP equation with the Yukawa potential can be obtained by inserting Eq. (22) into Eq. (21) as

$$\left[\frac{d^2}{dr^2} - 4a^2 J(J+1)\frac{e^{-2ar}}{(1-e^{-2ar})^2} + 4a^2U_0^2 \frac{e^{-4ar}}{(1-e^{-2ar})^2} + 4aEU_0 \frac{e^{-2ar}}{1-e^{-2ar}} - \varepsilon^2\right] F_{nJ}(r) = 0, \tag{24}$$

where $\varepsilon^2 = m^2 - E^2$. In the next section, we will introduce the generalized parametric NU method so that we can find solutions to the above equation.

## 3. Parametric NU Method

This powerful mathematical tool solves second order differential equations. Let us consider the following differential equation [40,41,42]

$$\psi_n''(s) + \frac{\tilde{\tau}(s)}{\sigma(s)}\psi_n'(s) + \frac{\tilde{\sigma}(s)}{\sigma^2(s)}\psi_n(s) = 0, \tag{25}$$

where $\sigma(s)$ and $\tilde{\sigma}(s)$ are polynomials, at most of second degree, and $\tilde{\tau}(s)$ is a first-degree polynomial. The application of the NU method can be made simpler and direct without need to check the validity of solution. We present a shortcut for the method. So, at first we write the general form of the Schrödinger-like equation (25) in a more general form as

$$\psi_n''(s)+\left(\frac{c_1-c_2 s}{s(1-c_3 s)}\right)\psi_n'(s)+\left(\frac{-p_2 s^2+p_1 s-p_0}{s^2(1-c_3 s)^2}\right)\psi_n(s)=0, \tag{26}$$

satisfying the wave functions

$$\psi_n(s)=\phi(s)y_n(s). \tag{27}$$

Comparing (26) with its counterpart (25), we obtain the following identifications:

$$\tilde{\tau}(s)=c_1-c_2 s,\quad \sigma(s)=s(1-c_3 s),\quad \tilde{\sigma}(s)=-p_2 s^2+p_1 s-p_0, \tag{28}$$

(1) For the given root $k_1$ and the function $\pi_1(s)$ :

$$k=-(c_7+2c_3c_8)-2\sqrt{c_8c_9},\quad \pi(s)=c_4+\sqrt{c_8}-\left(\sqrt{c_9}+c_3\sqrt{c_8}-c_5\right)s,$$

we follow the NU method [40] to obtain the energy equation [41,42]

$$nc_2-(2n+1)c_5+(2n+1)\left(\sqrt{c_9}+c_3\sqrt{c_8}\right)+n(n-1)c_3+c_7+2c_3c_8+2\sqrt{c_8c_9}=0, \tag{29}$$

and the wave functions

$$\rho(s)=s^{c_{10}}(1-c_3 s)^{c_{11}},\quad \phi(s)=s^{c_{12}}(1-c_3 s)^{c_{13}},\ c_{12}>0,\ c_{13}>0,$$

$$y_n(s)=P_n^{(c_{10},c_{11})}(1-2c_3 s),\ c_{10}>-1,\ c_{11}>-1,$$

$$\psi_{n\kappa}(s)=N_{n\kappa}s^{c_{12}}(1-c_3 s)^{c_{13}}P_n^{(c_{10},c_{11})}(1-2c_3 s). \tag{30}$$

where $P_n^{(\mu,\nu)}(x),\ \mu>-1,\ \nu>-1$, and $x\in[-1,1]$ are Jacobi polynomials with the constants are

$$c_4=\frac{1}{2}(1-c_1),\qquad c_5=\frac{1}{2}(c_2-2c_3),$$

$$c_6=c_5^2+p_2;\qquad c_7=2c_4c_5-p_1,$$

$$c_8=c_4^2+p_0,\qquad c_9=c_3(c_7+c_3c_8)+c_6,$$

$$c_{10}=2\sqrt{c_8}>-1,\qquad c_{11}=\frac{2}{c_3}\sqrt{c_9}>-1,\ c_3\neq 0,$$

$$c_{12}=c_4+\sqrt{c_8}>0,\qquad c_{13}=-c_4+\frac{1}{c_3}(\sqrt{c_9}-c_5)>0,\ c_3\neq 0, \tag{31}$$

where $c_{12}>0,\ c_{13}>0$ and $s\in[0,1/c_3],\ c_3\neq 0$.

In the rather more special case of $c_3=0$, the wave function (15) becomes

$$\lim_{c_3\to 0}P_n^{(c_{10},c_{11})}(1-2c_3 s)=L_n^{c_{10}}\left(2\sqrt{c_9}s\right),\ \lim_{c_3\to 0}(1-c_3 s)^{c_{13}}=e^{-(\sqrt{c_9}-c_5)s},$$

$$\psi(s)=Ns^{c_{12}}e^{-(\sqrt{c_9}-c_5)s}L_n^{c_{10}}(2\sqrt{c_9}s). \tag{32}$$

(2) For the given root $k_2$ and the function $\pi_2(s)$:

$k=-(c_7+2c_3c_8)+2\sqrt{c_8c_9},\ \ \pi(s)=c_4-\sqrt{c_8}-\left(\sqrt{c_9}-c_3\sqrt{c_8}-c_5\right)s,$

we follow the NU method [40] to obtain the energy equation

$$nc_2-(2n+1)c_5+(2n+1)\left(\sqrt{c_9}-c_3\sqrt{c_8}\right)+n(n-1)c_3+c_7+2c_3c_8-2\sqrt{c_8c_9}=0, \tag{33}$$

and the wave functions

$$\rho(s)=s^{\tilde{c}_{10}}(1-c_3s)^{\tilde{c}_{11}},\quad \phi(s)=s^{\tilde{c}_{12}}(1-c_3s)^{\tilde{c}_{13}},\ \tilde{c}_{12}>0,\ \tilde{c}_{13}>0,$$

$$y_n(s)=P_n^{(\tilde{c}_{10},\tilde{c}_{11})}(1-2\tilde{c}_3s),\ \tilde{c}_{10}>-1,\ \tilde{c}_{11}>-1,$$

$$\psi_{n\kappa}(s)=N_{n\kappa}s^{\tilde{c}_{12}}(1-c_3s)^{\tilde{c}_{13}}P_n^{(\tilde{c}_{10},\tilde{c}_{11})}(1-2c_3s), \tag{34}$$

where

$$\tilde{c}_{10}=-2\sqrt{c_8},\qquad \tilde{c}_{11}=\frac{2}{c_3}\sqrt{c_9},\ c_3\neq 0,$$

$$\tilde{c}_{12}=c_4-\sqrt{c_8}>0,\qquad \tilde{c}_{13}=-c_4+\frac{1}{c_3}(\sqrt{c_9}-c_5)>0,\ c_3\neq 0. \tag{35}$$

**4. Solution of the DKP-Yukawa problem**

Here we want to solve Eq. (24) in the context of the parametric generalization of the NU method. At first we introduce the change of variables $s=e^{-2ar}$ which maps the interval $(0,\infty)$ into $(0,1)$, to rewrite it as follows:

$$\frac{d^2F_{nJ}(s)}{ds^2}+\frac{1-s}{s(1-s)}\frac{dF_{nJ}(s)}{ds}+\frac{1}{s^2(1-s)^2}\Big[-J(J+1)s+U_0^2s^2$$
$$-\frac{E_{nJ}U_0}{a}s(1-s)-\frac{\varepsilon^2}{4a^2}\Big]F_{nJ}(s)=0. \tag{36}$$

Next, comparing Eq. (36) with its counterpart Eq. (26) enables us to find parametric coefficients $c_i$ ($i=1,2,3$) and analytical expressions $p_j$ ($j=1,2,3$) as follows

$$c_1=1,\qquad p_2=-U_0^2+\frac{E_{nJ}U_0}{a}+\frac{\varepsilon^2}{4a^2},$$

$$c_2=1,\qquad p_1=-J(J+1)+\frac{E_{nJ}U_0}{a}+\frac{2\varepsilon^2}{4a^2},$$

$$c_3=1,\qquad p_0=\frac{\varepsilon^2}{4a^2}. \tag{37}$$

The remaining values of coefficients $c_i$ ($i=4,5,...,9$) and $\tilde{c}_k$ ($k=10,...,13$) are found from relations (31) and (35). All values of these coefficients, i.e., $c_i$ ($i=4,5,...,9$) together with $\tilde{c}_k$ ($k=10,...,13$) are displayed in table 1. By using Eq. (33), we can obtain, in closed form, the energy eigenvalue equation as

$$\left[n+\frac{1}{2}+\left(\sqrt{\left(J+\frac{1}{2}\right)^2-U_0^2}-\sqrt{\frac{m^2-E_{nJ}^2}{4a^2}}\right)\right]^2=\frac{m^2}{4a^2}-\left(\frac{E_{nJ}}{2a}-U_0\right)^2. \qquad (38)$$

To find the energy levels we solve Eq. (38) numerically considering the following values of the parameters as $m=938.0\ MeV$, $U_0=67.54\,MeV$ [24] and screening parameter $a=(0.005,0.015)\ fm^{-1}$ [39]. These numerical results are displayed in table 2.

To find corresponding wave functions, referring to table 1 and relation (34), we get

$$F_{nJ}(s)=N_{nJ}s^{-\sqrt{\frac{\varepsilon^2}{4a^2}}}(1-s)^{\frac{1}{2}+\sqrt{\left(J+\frac{1}{2}\right)^2-U_0^2}}P_n^{\left(-2\sqrt{\frac{\varepsilon^2}{4a^2}},2\sqrt{\left(J+\frac{1}{2}\right)^2-U_0^2}\right)}(1-2s),\ s=e^{-2ar}, \qquad (39a)$$

or equivalently

$$F_{nJ}(r)=N_{nJ}e^{-\varepsilon r}\left(1-e^{-2ar}\right)^{\frac{1}{2}+\sqrt{\left(J+\frac{1}{2}\right)^2-U_0^2}}P_n^{\left(-\frac{\varepsilon}{a},2\sqrt{\left(J+\frac{1}{2}\right)^2-U_0^2}\right)}\left(1-2e^{-2ar}\right), \qquad (39b)$$

where $N_{nJ}$ is the normalization constant.

We can express the Jacobi polynomials in terms of the hypergemetric function [43,44]

$$P_n^{(\lambda,\eta)}(1-2x)=\frac{(\lambda+1)_n}{n!}\,{}_2F_1(-n,\lambda+\eta+1+n;\lambda+1;x), \qquad (40)$$

where $(y)_n=\Gamma(y+1)/\Gamma(y-n+1)$, is the Pochhammer's symbol.

The wave function (39b) satisfies the standard asymptotic analysis for $r\to 0$ and $r\to\infty$.

Further, we can take into account the following hypergeometric property [43,44]

$$\frac{d}{dx}\left({}_2F_1(b,c;d;x)\right)=\left(\frac{bc}{d}\right){}_2F_1(b+1,c+1;d+1;x), \qquad (41)$$

Thereby, we seek to find the other spinor components from Eqs. (17a), (17b) and (17c) as

$$G(r)=N_{nJ}\frac{(1-\varepsilon/\alpha)_n}{n!}\frac{1}{m}\left(E+U_0\frac{e^{-ar}}{r}\right)e^{-\varepsilon r}\left(1-e^{-2ar}\right)^{\frac{1}{2}+\sqrt{\left(J+\frac{1}{2}\right)^2-U_0^2}}$$

$$\times{}_2F_1\left(-n,-\frac{\varepsilon}{a}+2\sqrt{\left(J+\frac{1}{2}\right)^2-U_0^2}+1+n;-\frac{\varepsilon}{a}+1;e^{-2ar}\right), \tag{42a}$$

$$H_1(r)=-\frac{\alpha_J}{m}\left(\frac{1}{2}+\sqrt{\left(J+\frac{1}{2}\right)^2-U_0^2}\,\frac{2ae^{-2ar}}{\left(1-e^{-2ar}\right)}-\varepsilon-\frac{J+1}{r}\right)F(r)$$

$$+N_{nJ}\frac{(1-\varepsilon/\alpha)_n}{n!}\frac{\alpha_J}{m}\,{}_2F_1\left(-n+1,-\frac{\varepsilon}{a}+2\sqrt{\left(J+\frac{1}{2}\right)^2-U_0^2}+n+2;-\frac{\varepsilon}{a}+2;e^{-2ar}\right)$$

$$\times\frac{(2\alpha n)\left(2a\sqrt{\left(J+\frac{1}{2}\right)^2-U_0^2}+a+an-\varepsilon\right)}{(\varepsilon-a)}e^{-2ar}, \tag{42b}$$

$$H_{-1}(r)=-\frac{\varsigma_J}{m}\left(\frac{1}{2}+\sqrt{\left(J+\frac{1}{2}\right)^2-U_0^2}\,\frac{2ae^{-2ar}}{\left(1-e^{-2ar}\right)}-\varepsilon+\frac{J}{r}\right)F(r)$$

$$+N_{nJ}\frac{(1-\varepsilon/\alpha)_n}{n!}\frac{\varsigma_J}{m}\,{}_2F_1\left(-n+1,-\frac{\varepsilon}{a}+2\sqrt{\left(J+\frac{1}{2}\right)^2-U_0^2}+n+2;-\frac{\varepsilon}{a}+2;e^{-2ar}\right)$$

$$\times\frac{(2\alpha n)\left(2a\sqrt{\left(J+\frac{1}{2}\right)^2-U_0^2}+a+an-\varepsilon\right)}{(\varepsilon-a)}e^{-2ar}, \tag{42c}$$

## 4. $J=0$ Case

We consider the special case when the total angular momentum $J=0$ (s-wave). Thus, from Eq. (38) we obtain the following exact energy equation:

$$\left[n+\frac{1}{2}+\left(\sqrt{\frac{1}{4}-U_0^2}-\frac{1}{2\alpha}\sqrt{m^2-E_{n0}^2}\right)\right]^2=\frac{m^2}{4a^2}-\left(\frac{E_{n0}}{2a}-U_0\right)^2. \tag{43}$$

where $|m|>(E_{n0}-2\alpha U_0)$, $|m|>E_{n0}$ and $1>4U_0$. In addition, from Eqs. (39b) and (42), we also find the exact spinor components of the wave function as

$$G(r)=N_{n0}\frac{\left(1-\varepsilon/\alpha\right)_n}{n!}\frac{1}{m}\left(E+U_0\frac{e^{-ar}}{r}\right)e^{-\varepsilon r}\left(1-e^{-2ar}\right)^{\frac{1}{2}+\sqrt{\frac{1}{4}-U_0^2}}$$

$$\times{}_2F_1\left(-n,-\frac{\varepsilon}{a}+2\sqrt{\frac{1}{4}-U_0^2}+1+n;-\frac{\varepsilon}{a}+1;e^{-2ar}\right), \tag{44a}$$

$$H_1(r)=-\frac{\alpha_0}{m}\left(\frac{1}{2}+\sqrt{\frac{1}{4}-U_0^2}\frac{2ae^{-2ar}}{\left(1-e^{-2ar}\right)}-\varepsilon-\frac{1}{r}\right)F(r)$$

$$+N_{n0}\frac{\left(1-\varepsilon/\alpha\right)_n}{n!}\frac{\alpha_0}{m}{}_2F_1\left(-n+1,-\frac{\varepsilon}{a}+2\sqrt{\frac{1}{4}-U_0^2}+n+2;-\frac{\varepsilon}{a}+2;e^{-2ar}\right)$$

$$\times\frac{\left(2\alpha n\right)\left(2a\sqrt{\frac{1}{4}-U_0^2}+a+an-\varepsilon\right)}{\left(\varepsilon-a\right)}e^{-2ar}, \tag{44b}$$

$$H_{-1}(r)=-\frac{\varsigma_0}{m}\left(\frac{1}{2}+\sqrt{\frac{1}{4}-U_0^2}\frac{2ae^{-2ar}}{\left(1-e^{-2ar}\right)}-\varepsilon\right)F(r)$$

$$+N_{n0}\frac{\left(1-\varepsilon/\alpha\right)_n}{n!}\frac{\varsigma_0}{m}{}_2F_1\left(-n+1,-\frac{\varepsilon}{a}+2\sqrt{\frac{1}{4}-U_0^2}+n+2;-\frac{\varepsilon}{a}+2;e^{-2ar}\right)$$

$$\times\frac{\left(2\alpha n\right)\left(2a\sqrt{\frac{1}{4}-U_0^2}+a+an-\varepsilon\right)}{\left(\varepsilon-a\right)}e^{-2ar}. \tag{44c}$$

## 5. Conclusion

In this paper, we solved the Duffin–Kemmer–Petiau equation under a vector Yukawa potential. We used the conventional approximation scheme to the total angular momentum centrifugal term. The parametric NU method is used to obtain the energy eigenvalues and the corresponding spinor components of the eigenfunction. The explicit forms of the spinor components of wave function were calculated. In addition, some numerical results for bound state energies are given in table 2 various values of $(n,J)$ and screening parameter $a$.

**Acknowledgments**

S. M. Ikhdair acknowledges the partial support provided by the Scientific and Technological Research Council of Turkey.

**References**

[1] G. Petiau, Acad. R. Belg., A. Sci. Mém. Collect. **16** (2) (1936) 1.

[2] G. Petiau, Ph. D Thesis, University of Paris, 1936; Acad. R. Belg. Cl. Sci. Mém. Collect. 8, 16 (1936).

[3] N. Kemmer, Proc. R. Soc. A **166** (1938) 127.

[4] R. J. Duffin, Phys. Rev. **54** (1938) 1114.

[5] N. Kemmer, Proc. R. Soc. A **173** (1939) 91.

[6] B. C. Clark et al., Phys. Rev. Lett. **55** (1985) 592.

[7] G. Kalbermann, Phys. Rev. C **34** (1986) 2240.

[8] R .E. Kozack et al., Phys. Rev. C **37** (1988) 2898.

[9] R. E. Kozack, Phys. Rev. C **40** (1989) 2181.

[10] V. K. Mishra et al., Phys. Rev. C **43** (1991) 801.

[11] Y. Nedjadi and R. C. Barrett, J. Phys. G **19** (1993) 87.

[12] B. C. Clark et al., Phys. Lett. B **427** (1998) 231.

[13] M. Nowakowski, Phys. Lett. A **244** (1998) 329.

[14] J. T. Lunardi, B. M. Pimentel, R. G. Teixeira and J. S. Valverde, Phys. Lett. A **268** (2000) 165.

[15] V. Ya. Fainberg and B. M. Pimentel, Theor. Math. Phys. **124** (200) 1234.

[16] V. Ya. Fainberg and B. M. Pimentel, Phys. Lett. A **271** (2000) 16.

[17] B. Boutabia-Chéraitia and T. Boudjedaa, Phys. Lett. A **338** (2005) 97.

[18] D. A. Kulikov, R.S. Tutik and A.F. Yaroshenko, Mod. Phys. Lett. A **20** (2005) 43.

[19] F. Yaşuk, C. Berkdemir, A. Berkdemir and C. Önem, Phys. Scr. **71** (2005) 340.

[20] I. Boztosun, M. Karakoç, F.Yaşuk and A. Durmuş, J. Math. Phys. **47** (2006) 062301.

[21] Y. Kasri and L. Chetouani, Int. J. Theor. Phys. **47** (2008) 2249.

[22] A. S. de Castro, J. Phys. A **44** (2011) 035201.

[23] Y. Chargui, A. Trabelsia and L. Chetouani Phys. Lett. A **374** (2010) 2907.

[24]. M. Hamzavi and S. M. Ikhdair, Few Body Syst. (2012) (at press).
[25]. J. McEnnan, L. Kissel and R. H. Pratt, Phys. Rev. A **13** (1976) 532.
[26]. C. H. Mehta and S. H. Patil, Phys. Rev. A **17** (1978) 34.
[27]. R. Dutt and Y. P. Varshni, Z. Phys. A **313** (1983) 143; ibid. Z. Phys. D **2** (1986) 207; C. S. Lai and M. P. Madan, ibid. **316** (1984) 131.
[28]. S. M. Ikhdair and R. Sever, Int. J. Mod. Phys. A **21** (2006) 6465.
[29]. T. Imbo, A. Pagnamenta and U. Sukhatme, Phys. Lett. A **105** (1984) 183.
[30]. S. M. Ikhdair and R. Sever, J. Mol. Struct.: THEOCHEM **809** (2007) 103.
[31]. S. M. Ikhdair and R. Sever, J. Math. Chem. **41** (4) (2007) 329.
[32]. S. M. Ikhdair and R. Sever, J. Math. Chem. **41** (4) (2007) 343.
[33]. M. Karakoc and I. Boztosun, Int. J. Mod. Phys. E **15** (6) (2006) 1253.
[34]. B. Gönül, K. Köksal and E. Bakır, Phys. Scr. **73** (2006) 279.
[35]. S. M. Ikhdair, Cent. Eur. J. Phys. **10** (2012) 361.
[36]. E. Z. Liverts, E. G. Drukarev, R. Krivec and V. B. Mandelzweig, Few Body Syst. **44** (2008) 367.
[37]. R. L. Greene and C. Aldrich, Phys. Rev. A **14** (1976) 2363; M. R. Setare and S. Haidari, Phys. Scr. **81** (2010) 065201.
[38]. S. M. Ikhdair and R. Sever, J. Phys. A: Math. Theor. 44 (2011) 345301.
[39]. M. Hamzavi, S. M. Ikhdair and M. Solaimani, Int. J. Mod. Phys. E **21** (2012) 1250016.
[40] A. F. Nikiforov and V. B. Uvarov, Special Functions of Mathematical Physics, (Birkhausr, Berlin, 1988).
[41] S. M. Ikhdair, Int. J. Mod. Phys. C **20** (10) (2009) 1563.
[42] S. M. Ikhdair, Eur. Phys. J. A **39** (2009) 307.
[43] I. S. Gradsteyn and I. M. Ryzhik, Tables of Integrals, Series and Products, Academic Press, New York, 1994.
[44] S. M. Ikhdair and M. Hamzavi, Approximate Dirac solutions of complex PT-symmetric Pöschl-Teller potential in view of spin and pseudo-spin symmetries, Phys. Scr. (2012).

**Table 1.** The specific values for the parametric constants necessary for the energy eigenvalues and eigenfunctions

| Constant | Analytic value |
|---|---|
| $c_4$ | $0$ |
| $c_5$ | $-\frac{1}{2}$ |
| $c_6$ | $\frac{1}{4}-U_0^2+\frac{E_{nJ}U_0}{a}+\frac{\varepsilon^2}{4a^2}$ |
| $c_7$ | $J(J+1)-\frac{E_{nJ}U_0}{a}-\frac{2\varepsilon^2}{4a^2}$ |
| $c_8$ | $\frac{\varepsilon^2}{4a^2}$ |
| $c_9$ | $\left(J+\frac{1}{2}\right)^2-U_0^2$ |
| $\tilde{c}_{10}$ | $-2\sqrt{\frac{\varepsilon^2}{4a^2}}$ |
| $\tilde{c}_{11}$ | $2\sqrt{\left(J+\frac{1}{2}\right)^2-U_0^2}$ |
| $\tilde{c}_{12}$ | $-\sqrt{\frac{\varepsilon^2}{4a^2}}$ |
| $\tilde{c}_{13}$ | $\frac{1}{2}+\sqrt{\left(J+\frac{1}{2}\right)^2-U_0^2}$ |

**Table 2:** Bound state eigenvalues $E_{nJ}\,(MeV)$ of the DKP equation under a vector Yukawa potential for various $n$ , $J$ and screening parameter $a$ .

| $n$ | $J$ | $E_{nJ}$ | |
|---|---|---|---|
| | | $a = 0.005\ fm^{-1}$ | $a = 0.015\ fm^{-1}$ |
| 0 | 0 | -871.4020165 | -870.7176063 |
| | 1 | -923.6139661 | -922.9223014 |
| | 2 | -931.4867386 | -930.7744718 |
| | 3 | -934.1933978 | -933.4522010 |
| | 4 | -935.4372396 | -934.6588482 |
| | 5 | -936.1084061 | -935.2845563 |
| 1 | 0 | -922.1585762 | -921.4679150 |
| | 1 | -931.4201148 | -930.7082258 |
| | 2 | -934.1816502 | -933.4406736 |
| | 3 | -935.4339333 | -934.6556955 |
| | 4 | -936.1071998 | -935.2834677 |
| | 5 | -936.5088793 | -935.6314014 |
| 2 | 0 | -930.9974392 | -930.2877285 |
| | 1 | -934.1535673 | -933.4131126 |
| | 2 | -935.4279337 | -934.6499738 |
| | 3 | -936.1052972 | -935.2817504 |
| | 4 | -936.5081276 | -935.6307868 |
| | 5 | -936.7655234 | -935.8261476 |
| 3 | 0 | -933.9779358 | -933.2405413 |
| | 1 | -935.4136146 | -934.6363132 |
| | 2 | -936.1018465 | -935.2786346 |
| | 3 | -936.5069413 | -935.6298173 |
| | 4 | -936.7650274 | -935.8258091 |
| | 5 | -936.9380488 | -935.9285043 |
| 4 | 0 | -935.3248226 | -934.5514102 |
| | 1 | -936.0936195 | -935.2712010 |
| | 2 | -936.5047911 | -935.6280589 |
| | 3 | -936.7642455 | -935.8252750 |
| | 4 | -936.9377089 | -935.9283417 |
| | 5 | -937.0579382 | -935.9699520 |
| 5 | 0 | -936.0428896 | -935.2251697 |
| | 1 | -936.4996693 | -935.6238636 |
| | 2 | -936.7628287 | -935.8243063 |
| | 3 | -936.9371726 | -935.9280848 |
| | 4 | -937.0576990 | -935.9699094 |
| | 5 | -937.1430640 | -935.2845563 |

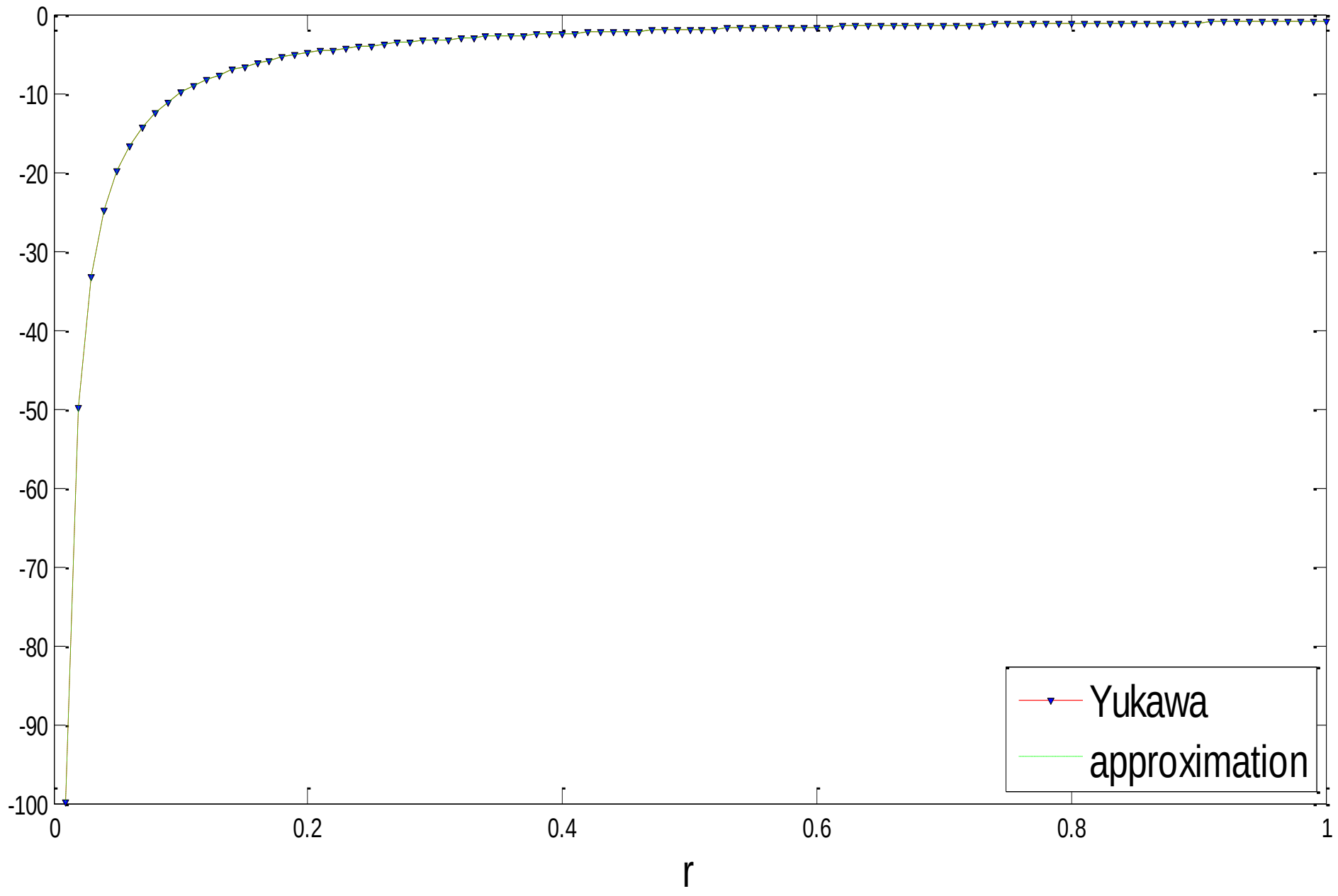

**Figure 1:** Yukawa potential (1) (red curve) and its approximation (23) (green curve).